\documentstyle[12pt,psfig]{article}
\advance\textheight by\topskip

\begin{document}

\title{ RATAN-600 -- VLA -- BTA-6\,m (``BIG TRIO'') PROJECT:
MULTICOLOUR STUDYING OF DISTANT RADIO GALAXIES
}

\author{ Yu.N. Parijskij{$^{(3)}$}, W.M. Goss{$^{(2)}$},
         A.I.Kopylov{$^{(3)}$}, N.S. Soboleva{$^{(1)}$},
\and     A.V. Temirova{$^{(1)}$}, O.V. Verkhodanov{$^{(3)}$},
         O.P. Zhelenkova{$^{(3)}$}
\and{\small\it (1) St.Petersburg Branch of Special Astrophysical Observatory,
    }
\and{\small\it Pulkovo, St.Petersburg, Russia }
\and{\small\it e-mail: par,sns,tem@fsao.spb.su }
\and{\small\it (2) National Radio Astronomy Observatory,
                P.O. Box 0, Socorro,NM, 87801, USA }
\and{\small\it e-mail: mgoss@aoc.nrao.edu }
\and{\small\it (3) Special Astrophysical Observatory of Russian Academy of
         Sciences, }
\and{\small\it Nizhnij Arkhyz, Karachai-Cherkessia, 357147, Russia}
\and{\small\it e-mail: akop,vo,zhe@sao.ru }
}
\date{}
\maketitle

\normalsize
\begin{abstract}
Powerful radio galaxies belong to the population of massive stellar
systems that can be picked at high redshifts.
With measured redshift, morphology and age determination
these objects can help greatly in understanding of
the way of formation and evolution of giant
elliptical galaxies with massive Black Holes in the Early Universe.

We presents some recent results of study of the sample of 105 faint
steep spectrum radio sources from RATAN-600 RC catalogue.
All the objects has been mapped with the VLA and optical identification
have been carried out with the 6\,m BTA telescope.
Using multicolour CCD photometry in $B,V,R_c,I_c$ bands
photometric redshift and age of stellar population
are estimated for a subsample of 50 FRII and CSS radio galaxies.
RC J0105+0501 is a very probable candidate for radio galaxy
at $z\approx 3.5$.

Distribution of objects on colour--colour and magnitude--redshift
diagrams shows a rough agreement with expectations of theoretical
models of SED evolution for giant elliptical galaxies.
On the [$m_R - z_{phot}$] diagram two populations of faint
($m_R = 22^m-24^m$) radio galaxies are possibly revealed.
The first population at $z\approx 1$ consists of mainly old (a half with
age $\geq 5 Gyr$) and large ($LAS\sim 20''$) objects, the second one
at z = 1.5 -- 3.5 includes both younger ($\leq 3 Gyr$) and smaller
($LAS\sim 5''$) objects.

There are 12 more faint or still undetected ($m_R\geq 24^m.5$)
sources in the sample which may presumably be a very distant
radio galaxies ($z>3.5$) or intermediate redshift old or dusty ones.

\end{abstract}

\section {Introduction}
\hspace*{\parindent}
Radio galaxies of FRII type (FRII RGs) belong to the most powerful
radio objects of the Universe and at
least at low redshifts they are connected with the most massive galaxies
with the extreme massive ($\sim 10^9 M_\odot$) and energetic central engines.
Evolution of this population and giant Black Hole inside is not well
understood yet.
Many groups try to check by observations very different scenarios
of stellar population evolution existing in literature.

It was realised recently
that cm-wavelength 10--50 mJy flux density range is the very interesting:
FRII objects dominate here and at the same time they are
rather bright yet to be observed optically.
Suggested approach optimize the solution of the radio selection
of the objects which were born at very high redshifts (HZ).
The ``Big Trio'' project incorporates opportunities of three
large telescopes (RATAN-600, VLA and BTA-6\,m) to
realize a selection of the powerful
radio galaxies between extremely deep ``pencil beam''
very small field surveys (HST, VLA), were fields are smaller
than mean distance between HZ FRII RGs, and all sky surveys
(NVSS, FIRST, SLOAN),
were objects can not be explored deeply ($R>23$) optically in the most
interesting radio flux density range.

\section {The sample of RC steep spectrum objects}
\hspace*{\parindent}
The ``Big Trio'' project belongs to the deepest ones in attempts
to select the powerful FRII radio galaxies
with ultra steep spectra (FRII USS RG) at limiting flux density level
below 3C, PKS, B2, GB samples of FRII USS objects.

RATAN-600 RC catalogue of 1145 radio sources in
$\approx$200 sq. degrees strip (at DEC$\approx 5^{\circ}$, $40'$ wide)
with limiting flux density $S_{3.9GHz}\approx 10mJy$
(Parijskij et al., 1991; 1992) was used to select
105 USS objects ($\alpha \geq 0.9$) by cross-identification with
a preliminary version (1988), kindly provided by Douglas prior to
publication, of
$0.365GHz$ Texas (UTRAO) catalogue (Douglas et al., 1996).
All selected objects were mapped by VLA to determine precise coordinates,
largest angular size (LAS) and radio morphology and optically
identified down to $m_R\leq 25^m$ by 6\,m Russian telescope
(Goss et al., 1992; Kopylov et al., 1995a,b; Fletcher et al., 1996;
Parijskij et al., 1996; 1998). Also direct imaging data of 22 objects
were obtained at (or near) a subarcsecond seeing with 2.56\,m Nordic
Optical Telescope at La Palma (Pursimo et al., 1999).
In Table 1 ranges and medians of main characteristics of objects of
our USS sample are given.

\begin{table}[h]
\caption{Sample of RC USS objects}
\begin{center}
\begin{tabular}{lll}\hline
Parameter & Range & Median\\
\hline
$S_{3.9GHz}$ & $15-350$ mJy & 67 mJy\\
$\alpha^{0.365}_{3.9}$ & $0.9-1.5$ & 1.0\\
LAS & $\leq0''.7-120''$ & $10''$\\
$m_R$ & $18^m-\geq25^m$ & $22^m.5$\\
\hline
\end{tabular}
\end{center}
\end{table}

There are 33 compact steep spectrum (CSS) objects in the sample
16 of which has $1''<LAS<4''$ and 17 are unresolved or barely
resolved ($LAS<1''$).
65 objects look like FRII and about 20 of them belong
presumably to the most distant generation of RGs.
16 objects were classified as quasars by their stellar appearance
on CCD images.

\section{BVRI-photometry}
\hspace*{\parindent}
The technique of multicolour photometry has became in the past few
years as the main method in selecting candidates for distant galaxies,
and the only approach at very large redshifts. Determination of
the age of HZ stellar systems may be the only way of estimation of
first galaxies formation redshift if star formation begins at redshifts
larger than $z$ of secondary ionization.
Direct observation of protogalaxies predicted by some recent computer
simulations are not possible. It was shown (Verkhodanov et al., 1999)
that namely BVRI colours are
sufficient for accurate estimation of $z$ and age in
the redshift range of 0.5--3.5.

We have implemented this approach
for 50 FRII and CSS radio galaxies of the RC catalogue that
had been observed in $B,V,R_c,I_c$ bands
during 1994--1998 with CCD camera on 6\,m telescope.
Our sample of USS radio galaxies is now the largest one
with a four-band optical photometry.
The data obtained were used to estimate colour photometric redshifts
($z_{phot}$) and ages of host galaxies by comparison with
two models of evolution of spectral energy distribution (SED)
(Fioc and Rocca-Volmerange, 1997; Poggianti, 1997).
Few typical cases were checked spectroscopically and reality of the colour
$z$ determination was confirmed (Dodonov et al., 1999).

\subsection {A high redshift radio galaxy RC J0105+0501}

\begin{figure*}[t]
\centerline{
\vbox{\psfig{figure=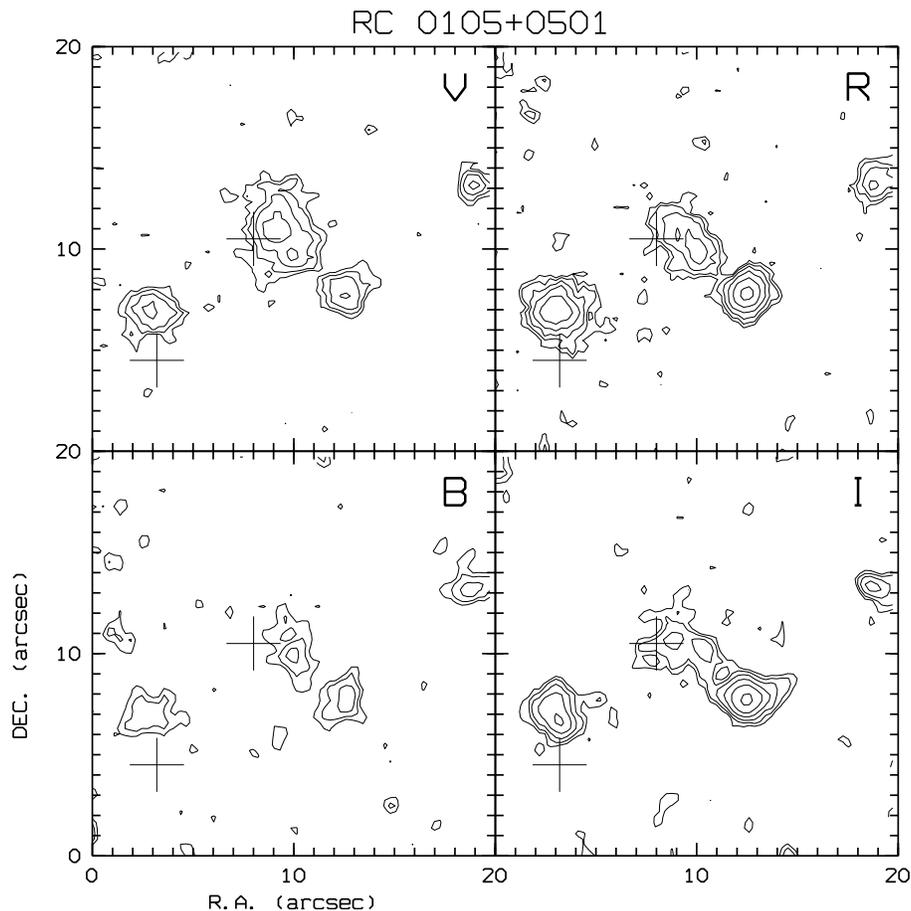,width=12.0cm,angle=-90,bbllx=40pt,bblly=83pt,bburx=542pt,bbury=585pt,clip=}}
}
\caption{\it
Contour maps in $B,V,R_c,I_c$ bands of RC J0105+0501.
Images with seeing of FWHM=$1''.4$ were obtained in August, 1998 with
exposures of 600, 400, 400 and 2$\times$400~s in $B$, $V$, $R_c$ and $I_c$,
respectively. Positions of two radio components are shown by pluses.
North is up and East is to the left.
}
\end{figure*}

\hspace*{\parindent}
As the best example of high redshift population of our sample
we presents $S_{3.9GHz}=33 mJy$ radio galaxy RC J0105+0501
($m_V=22^m.5$ complex object at the center of four boxes on Figure 1),
which shows the colour properties and the structure characteristic of
very distant powerful radio galaxies. In the $V$ band, the galaxy is
most extended and is by $1^m.5$ brighter than in the $B$ band, which is
interpreted almost unambiguously for the given class of objects as a
powerful Ly$\alpha$ line emission and continuum depression in the adjacent
region of shorter wavelengths. The negative colour index $V$--$R_c=-0.3$ and
the small index $R_c$--$I_c$=0.4 are in agreement with this interpretation
of the data. The redshift is estimated to be 3.4 $\pm$ 0.3.
(The first radio galaxy with $z>3$ was discovered by Lilly (1988), but
only about 20 objects of this kind (RG at $z>3$) have been found so far
by joint efforts of different groups.
The current ``champion'' has $z=5.19$ (Van Breugel et al., 1999).)

It can be seen that in $V$-band the host galaxy
is resolved into two components separated by $1''.7$.
(The two brighter neighbouring galaxies, $\approx 4''$ to SW and
$\approx 8''$ to SE, are likely to
have nothing to do with the radio galaxy since the colour redshift for them
is estimated to be about 1.)
The SW-component is reliably detected on the $B$ frame and may be an
active nucleus of a radio galaxy. The second component may then be either
a region of star formation induced by the jet or a gaseous cloud ionized
by the radiation from the active nucleus or a combination of both.
Other interpretations are also possible. For instance, the active nucleus
may be identified with the NE-component or even coincide with the radio
component without showing up in optics because of the strong absorption
by dust. The necessity of more detailed study of this very fascinating
object in the optical, IR and radio ranges for testing different hypotheses
on the physics of the processes occurring in this first generation stellar
system is evident.

\subsection{Colour -- Colour distribution}

\begin{figure*}[t]
\centerline{
\vbox{\psfig{figure=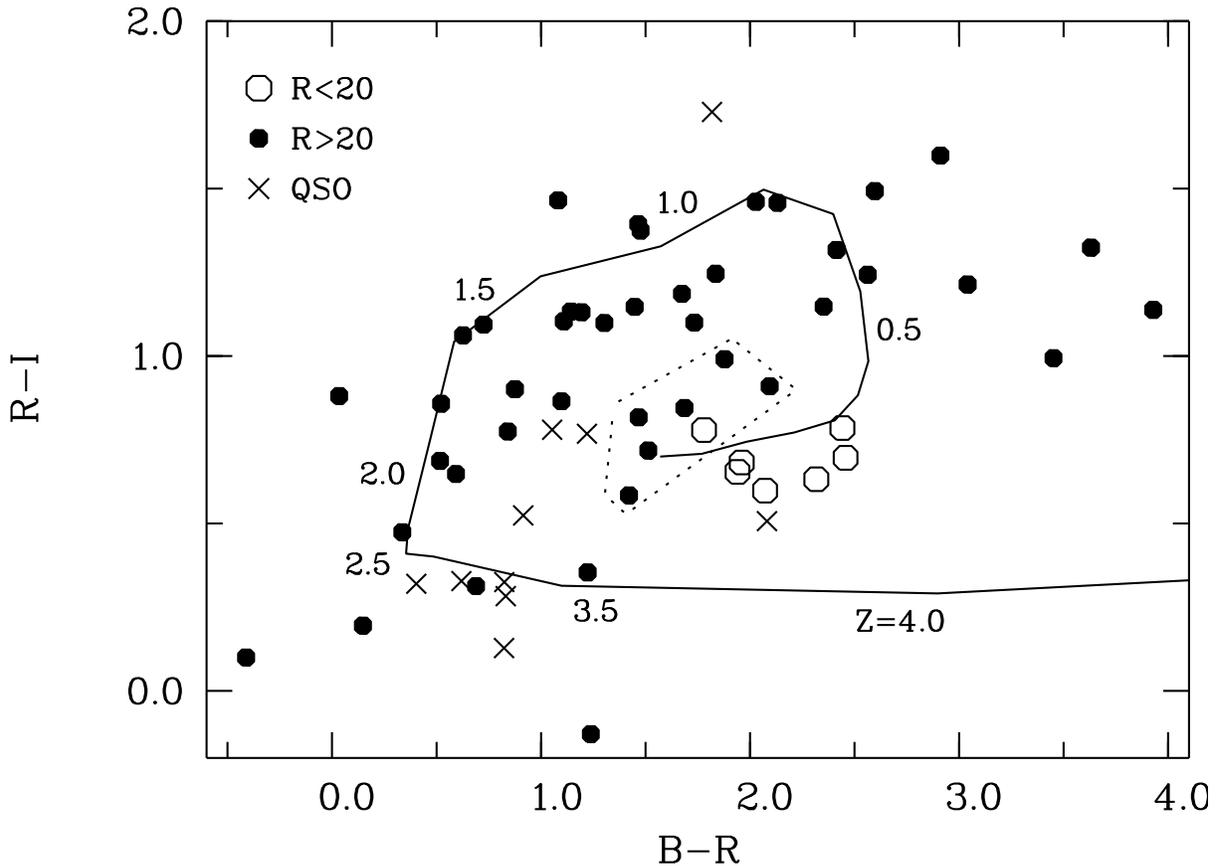,width=16.0cm,angle=-90,bbllx=127pt,bblly=79pt,bburx=524pt,bbury=628pt,clip=}}
}
\caption{\it
$B$--$R_c - R_c$--$I_c$ diagram for 50 RGs (objects with $m_R<20^m$ are
shown by open circles and with $m_R>20^m$ by closed circles)
and 10 quasars (crosses). Galaxies with probable strong contribution
of AGN light or young stellar population lie inside a dotted contour.
Evolutionary track (with redshifts marked) of PEGASE model of giant
elliptical galaxy with $z_{form}$=15 and present day age of 16 Gyr
is shown for $H_0$=50 km/s/Mpc and $\Omega_0$=0.1 cosmology.
}
\end{figure*}

\hspace*{\parindent}
A two-colour distribution of RC-objects so far observed
in comparison with PEGASE evolutionary model of giant
elliptical galaxy with a formation redshift $z_{form}$=15
and present day age of 16 Gyr is given on Figure 2.
There is a rough agreement between observations and a theoretical model
taking into account typical errors in colours of 0$^m$.2--0$^m$.3.
These are larger for 3 objects most red in $B$--$R_c$ colour which were
observed at the detection limit in $B$-band.
As for 3 objects most blue in $B$--$R_c$ a strong Ly$\alpha$ line emission
in $B$-band could be suspected (and indeed
conformed spectroscopically for one of them (Dodonov et al., 1999)).
A plausible explanation for several objects lying inside a dotted contour
on Figure 2 may consist in that a noticeable contribution of
the active nucleus of galaxy or a mixture of young and old stellar
populations is observed. In both cases a shift of $0^m.5-1^m.0$ in
upper-right direction may bring points to the location of main
stellar population of host galaxy. The same effect could explains
$R_c$--$I_c$ colours of group of points below the model
curve near $z\sim 1$. Alternatively, a formation redshift,
an underlying cosmological model or a stellar content of the SED model
could be varied. All these considerations are by no means
conclusive until spectroscopic measurement of redshifts will be done.
What could be affirmed now is that as a whole the sample of RC FRII
USS objects follows in a redshift range of 0.2--3.5
the predictions of PEGASE and, not shown on Figure 2,
Poggianti's  models of SED evolution of giant elliptical galaxies formed
at high redshifts.

\section{Magnitude -- Redshift  diagram}

\begin{figure*}[t]
\centerline{
\vbox{\psfig{figure=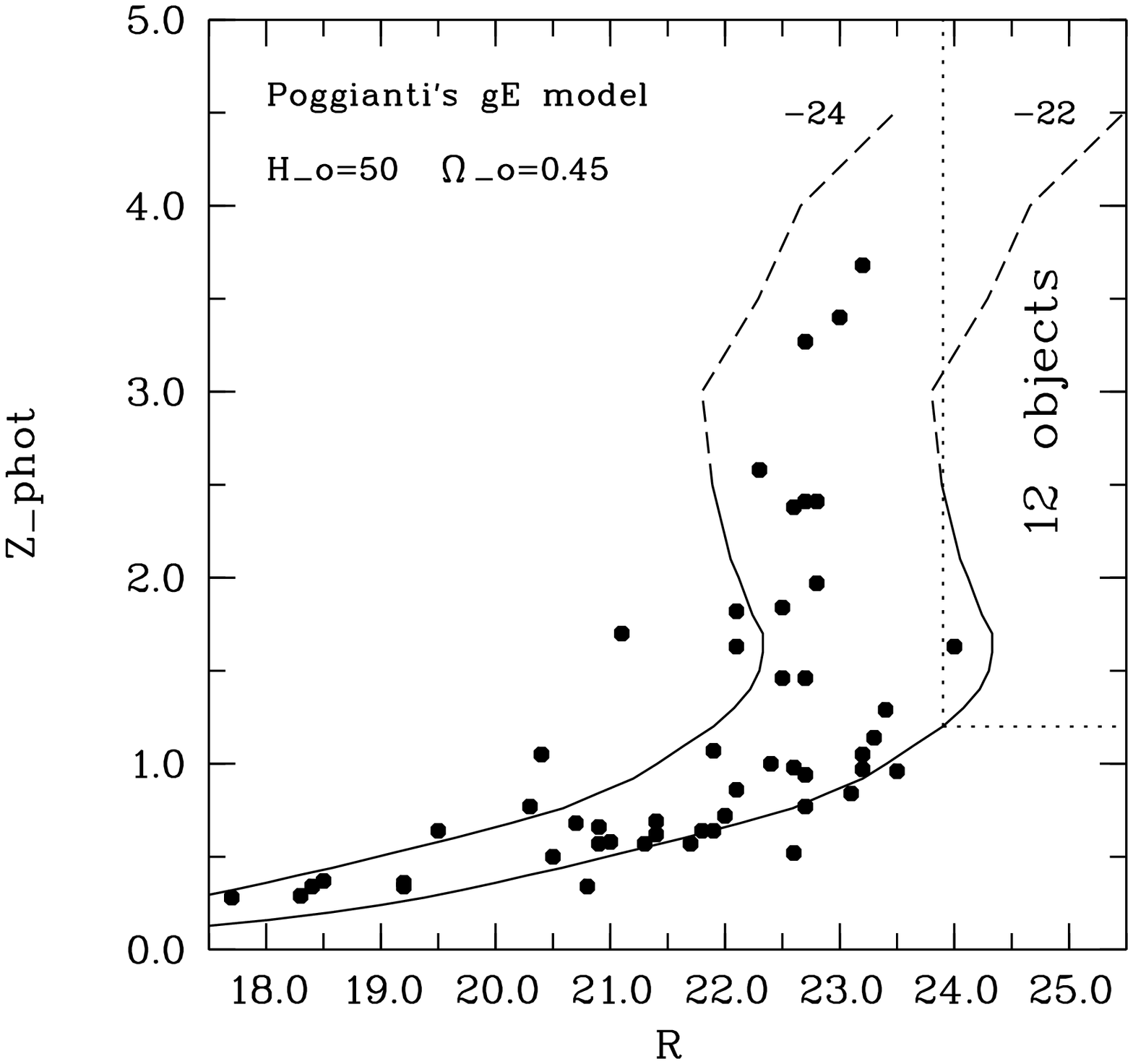,width=16.0cm,angle=-90,bbllx=72pt,bblly=99pt,bburx=542pt,bbury=599pt,clip=}}
}
\caption{\it
$m_R - z_{phot}$ diagram for 50 RGs.
Evolutionary tracks of Poggianti's model of elliptical galaxies of
$M_R=-24^m$ and $M_R=-22^m$ are shown by lines.
Inside a box shown by dotted line in upper-right corner 12 very faint or
still undetected objects should be located.
}
\end{figure*}

\hspace*{\parindent}
A model of a passively evolving giant elliptical
galaxy predicts quite definite magnitude dependence on redshift
if star formation in galaxy begins at a rather early time.
So a distribution of objects on, for example, a $m_R - z_{phot}$ diagram
can be considered as an indirect check of the accuracy of photometric
redshift estimates for the sample as a whole taking into account that
powerful radio sources live as a rule in luminous
($M_R\leq M^{\ast}\approx -22^m.5$) elliptical galaxies (if there are no
special problems with dust at high redshift, dispersion in $z_{sf}$ and
homogenity of the sample).

As can be seen on Figure 3 there is no great contradiction between
observed distribution and expected one for Poggianti's model in
an ``intermediate'', $\Omega_0 = 0.45$, cosmological model.
The same can be said for PEGASE model (not shown on Figure 3).
Unexpected to some extent is rather low mean redshift $<z_{phot}>=1.6$
for 27 objects with
magnitudes of $22^m\leq m_R\leq 24^m$. Formerly we have estimated
(Parijskij et al., 1998) that a mean redshift should be
of about 2 for such faint objects using 3 types of calibrations
(by $LAS$, $\alpha$ and $m_R$), based on a large data set from literature
on radio galaxies with measured $z\geq 1$.
But several effects could help in interpretation of
the difference.

Though redshifts, estimated by $B,V,R_c,I_c$, should be more precise
they may has its own systematics.
On average we have a fainter radio sources in RC USS sample
than in other ones but with lower median $\alpha$ (Table 1).
The latter property of the sample my be of greater importance for
a selection of HZ objects.
Indeed, the calibration by $\alpha$ had given the smallest of all
$<z>\approx1.5$.
A more sophisticated interpretation, which seems only speculative until
a completion of multicolour observations of remaining objects
and spectroscopic redshifts measurements, is consisted
in that on the $m_R - z_{phot}$ diagram two populations of faint
($m_R = 22^m-24^m$) radio galaxies are possibly revealed.
The first population at $z\approx 1$ consists of old (a half with age
$\geq 5 Gyr$) and large ($LAS\sim 20''$) objects, the second one at
z = 1.5 -- 3.5 includes both younger ($\leq 3 Gyr$) and smaller
($LAS\sim 5''$) objects. The low $z$ population may becoming to dominate
for flux level and typical $\alpha$ of RC USS catalogue, thus providing
an explanation of lower mean redshift of the whole sample than
expected from calibrations based on objects with higher flux level
and steeper spectra.

While all but one of our objects with $m_R < 22^m$ have multicolour data,
there are 17 ones with $22^m\leq m_R < 24^m$ and another 12 with
$m_R\geq 24^m$ which was not observed in $BVRI$ so far.
Most of the latter ones should probably be at $z\geq 1.5$,
and some may be a very distant radio galaxies ($z>3.5$)
or intermediate redshift very old or dusty ones.\\

\section{Discussion}
\hspace*{\parindent}
Multicolour photometry is of special importance for $z>2$, where simple
(one-band) $z_{phot}$ may give very large errors.

There is increasing interest in the very old distant stellar systems.

Looking at {\it age -- z}{\,} relation for our sample of RG, we see well
the expected trend for parent
galaxies to be younger at higher redshifts, and we believe,
that selection effects play a secondary role in this result.
We confirm also, that mean epoch of the parent galaxies formation may
be in the redshift range 10--20, but in several cases the multicolour ages
exceed the SCDM Universe age at the estimated redshift.

First case of that kind we have mentioned in (Parijskij et al, 1996a),
another two
cases appeared recently (Spinrad et al., 1997; Cowan et al., 1997).
These cases were used to estimate the role of the $\Lambda$-term
(Yoshii et al., 1998; Alcaniz and Lima, 1999).
Even more, it was suggested to use {\it age -- z}
relation  for reconstruction of the physical conditions in the very early
Universe (Saini et al., 1999 and Starobinsky, 1999).

It is well known, that errors of colour ages may be small enough
for young stellar population, but for old population it is not the case,
and, at some redshifts ranges, colours are not sensitive to age at all.

Dust reddening may imitate the old age of the galaxy. But, it can be shown,
that separation of these effects may be done, due to the different shape of
observable SED.

High importance of the age determination suggests, that all possible ways of
the improving ages estimation should be
used, and $FeII/MgII$ ratio is one of the suggested for distant objects
with emission lines (Yoshii et al., 1998). We hope to select most interesting
objects for future studies, including deep spectroscopy.

\hspace*{\parindent}

{\bf Acknowledgements}
This work has been partially supported by grants from the Russian
Foundation for Basic Research 95-02-03783, 96-02-16597, 99-02-17114 and
99-07-90334, the State Science and Technology Programme
``Astronomy'' (grants 2-296, 1.2.1.2 and 1.2.2.4), INTAS 97-1192
and VUZ-RAS AO145.
The Very Large Array is funded by the National Science Foundation,
under a cooperative agreement with Associated Universities, Inc.

{}

\end{document}